\documentstyle[11pt,aaspp4,psfig,flushrt,tighten]{article}
\newcommand{\be}{\begin{equation}} \newcommand{\ba}{\begin{eqnarray}}
\newcommand{\ee}{\end{equation}} \newcommand{\ea}{\end{eqnarray}}
\newcommand{\etal}{et al.\ } \def\gtsima{$\; \buildrel > \over \sim
\;$} \def\ltsima{$\; \buildrel < \over \sim \;$}
\def\gsim{\lower.5ex\hbox{\gtsima}}
\def\lsim{\lower.5ex\hbox{\ltsima}}
\def\simgt{\lower.5ex\hbox{\gtsima}}
\def\simlt{\lower.5ex\hbox{\ltsima}}
\def\simpr{\lower.5ex\hbox{\prosima}}  
 \def\ekin{\mbox{\cal E}_{\rm kin}}
\def\ekin{{\cal E}_{\rm kin}} 
 
 \def\msun{{M_\odot}}
 \def\eg{{\frenchspacing\it e.g. }}
\def\Eg{{\cal E}_{\rm g}}  
\def\CIII{C{\sc ~iii} } \def\CIV{C{\sc ~iv} } \def\SiIV{Si{\sc ~iv} }
\def\OVI{O{\sc ~vi} }

\begin{document}
\title{What can the Distribution of Intergalactic Metals Tell us About
  the History of Cosmological Enrichment?}

\author{Evan Scannapieco\altaffilmark{1}} 
\altaffiltext{1}{Kavli Institute for
Theoretical Physics, Kohn Hall, UC Santa Barbara, Santa Barbara, CA
93106}

\begin{abstract}

I study the relationship between the spatial distribution of
intergalactic metals and the masses and ejection energies of the
sources that produced them. Over a wide range of models, metal
enrichment is dominated by the smallest efficient sources,  as the
enriched volume scales roughly as $E^{3/5} \sim M^{3/5}$ while the
number density of sources goes as $\sim M^{-1}.$   In all cases,
the earliest sources have the biggest impact, because fixed
comoving distances correspond to smaller physical distances at higher
redshifts.  This means that most of the enriched volume is found
around rare peaks, and intergalactic metals are naturally highly
clustered. Furthermore, this clustering is so strong as to lead  to a
large overlap between individual bubbles.   Thus the typical radius of
enriched $z \sim 3$ regions should be interpreted as a constraint on
{\em groupings} of sources rather than the ejection radius of a
typical source.  Similarly, the clustering of enriched  regions should
be taken as a measurement of source bias rather than mass.
\end{abstract}


\section{Introduction}

It is now clear that the intergalactic metals detected at $z \sim 3$ 
have only  a secondary impact on  further structure formation.   Numerous
studies of the \CIV, \SiIV, \OVI  and \CIII content of the
intergalactic medium (IGM) have been carried out (\eg Songaila \&
Cowie 1996;  Aracil \etal 2004; Aguirre \etal 2005)  only to find that this
material is a small fraction of the metals produced (Pettini 1999);
there have been many detailed measurements of $z \lesssim 5$ galaxy
outflows  (\eg Pettini 2001; Schwartz \& Martin 2004),  but the intergalactic
metal distribution is observed to be roughly constant over this entire
range (Songaila 2001; Pettini \etal 2003); and metal ejection has been
incorporated into several simulations (\eg Thacker \etal 2002;
Springel \& Hernquist 2003; Cen \etal 2005),  which found that it has
a negligible impact on IGM cooling and the statistical properties of
the Lyman-alpha forest (Theuns \etal 2002; Bruscoli 2003).   In short,
the IGM metals that we see are not doing very much.

Yet as a {\em tracer} of the higher-redshift interplay between
galaxies and the IGM, intergalactic metals are unparalleled.  It is
indeed remarkable that this material has made its way from the centers
of stars into the lowest-density environments detectable (Schaye
\etal 2003; Aracil \etal 2004), with far-reaching implications.
The depletion of metals through outflows 
directly  impacts the galaxy mass-metallicity relation (\eg Dekel \&
Woo 2003; Tremonti \etal 2004); outflows suppress
the formation of nearby objects  (Scannapieco
\etal 2000;   Sigward \etal 2005); and the distribution of IGM
metals is closely linked to the evolution of the first
generation of stars (\eg Bromm 2003;  Scannapieco \etal 2003).

Extracting the details of each of these processes from observation, however,
requires an uncertain extrapolation from $z \sim 3$ to much higher 
redshifts.  One approach to this problem is to focus on deriving 
constraints from the IGM composition, which can be
related to the star formation history, initial mass function, 
and metallicity of the sources (\eg Aguirre \etal 2004; 
Qian \& Wasserburg 2005).  In this case the
primary complications are due to uncertainties in  abundances and 
the ionizing background.

A second method relies on the spatial distribution of intergalactic
metals. Regardless of the details of IGM enrichment, it is clear that
metals were formed in  the densest regions of space, regions that are
far more clustered than the overall matter distribution.  Furthermore,
this ``geometrical biasing'' is a systematic function of mass and
redshift (\eg Kaiser 1984), and thus the observed
large-scale clustering of metal absorbers encodes  information on the
scales of the  objects from which they were ejected.  Likewise, the
size of each enriched region is dependent on the energy at which the
metals were dispersed.

Thus recent measurements of the sizes of enriched regions  (\eg Rauch
\etal 2001), the metallicity as a function of  IGM density (\eg Schaye
\etal 2003),  the \CIV absorber correlation function  (Rauch \etal
1996; Pichon \etal 2003), and the galaxy-\CIV cross-correlation
function (\eg Adelberger 2003), are already providing useful
constraints on simulations of metal enrichment.  Yet such detailed
comparisons are too expensive to carry out over a large  range of
parameter space, and provide little intuition.  One is left wondering
if perhaps there might be some general rules of thumb that can be kept
in mind when interpreting observations or selecting simulation
parameters.

It is this issue that I take up in this {\em Letter}.  Adopting a
simplified model, I show that the sizes and
clustering of the enriched regions that we see should  {\em not} be
interpreted as ejection radii of individual sources or correlated with
masses at $z \sim 3,$ but that nevertheless these quantities have
natural interpretations related to the properties
of the sources of IGM enrichment.
Throughout, I assume a cold dark matter
cosmological model with parameters $h=0.65$, $\Omega_0$ = 0.3,
$\Omega_\Lambda$ = 0.7, $\Omega_b = 0.05$, $\sigma_8 = 0.87$, and
$n=1$, where $h$ is the Hubble constant in units of 100 km/s/Mpc,
$\Omega_0$, $\Omega_\Lambda$, and $\Omega_b$ are the
total matter, vacuum, and baryonic densities in units of the critical
density, $\sigma_8^2$ is the linear variance on the $8
h^{-1}{\rm Mpc}$ scale, and $n$ is the ``tilt'' of the primordial
power spectrum (\eg Spergel \etal 2003), with the transfer
function taken from Eisenstein \& Hu (1999).

The structure of this work is as follows.  In \S 2 I describe a 
general model of cosmological enrichment and apply it
in \S 3 to derive the clustering properties
and sizes of enriched regions under a wide range of 
assumptions.  I conclude with a short discussion in \S 4.

\section{Modeling Cosmological Enrichment}

As we are interested in making general statements, 
I adopt here an extremely simple approach.   All
outflows are taken to be pressure-driven 
spherical shells, expanding into the Hubble
flow (\eg Ostriker \& McKee 1988).
To keep this model as transparent as possible, I
do not  attempt to include secondary effects, such as cooling,
a stochastic star formation rate, external pressure,  or
the gravitational drag from the source halo 
(Madau \etal 2001; Scannapieco \etal 2002).
In this case the evolutionary equations are
\ba \dot{R_s}
&=& \frac{3 P_b}{\bar \rho R_s} - \frac{3}{R_s}(\dot{R_s} - HR_s)^2 -
\Omega_m  \frac{H^2 R_s}{2}, \nonumber \\ 
\dot{E_b} &=&  L(t) - 4 \pi R_s^2 \dot{R_s} P_b,
\label{eq:rs}
\ea
where the overdots represent time derivatives, the subscripts {\it s}
and {\it b} indicate shell and bubble quantities respectively, $R_s$
is the physical radius of the shell, $E_b$ is the internal energy of the
hot bubble gas, $P_b$ is the pressure of this gas, and $\bar \rho$ is
the mean IGM background density. I assume adiabatic expansion
with an index $\gamma=5/3$ such that $P_b=E_b/2\pi R_s^3.$ When outflows
slow down to the sound speed, the shell is likely to fragment.
At this point I let the region expand with the Hubble flow.

In this model, the evolution of the shell is completely determined by
the mechanical luminosity, 
\be
L(t) =  160 \, L_\odot  \, f_\star \,  f_w  \, {\cal N} \,  M_b 
\, \Theta(t_{\rm SN}-t),  
\label{eq:lum}
\ee 
where $f_\star$ is the fraction of gas converted into stars, $f_w$ is the
fraction of the supernova (SN) kinetic energy that is channeled into the galaxy
outflow, ${\cal N}$ is the number of SNe per solar mass of stars
formed (each assumed to explode with $10^{51}$ ergs of kinetic energy),
$M_b$ is the baryonic mass of the galaxy in units of solar
mass, and $t_{\rm SN} = 5 \times 10^7$ years.   
Following Scannapieco \etal (2003),
I define the product, $f_\star f_w \ekin {\cal N}$,
as the ``energy input per unit gas mass'' ${\cal E}_{\rm g}$.
Assuming $f_\star = 0.1,$  $f_w = 0.1,$ and 1 SN per 300 solar masses 
gives a fiducial estimate for $\Eg$ of $10^{-4.5}$,
although I vary this parameter over a wide range below.
Note that the choice of $t_{\rm SN}$ has no direct impact on 
the results.

By combining eqs.\ (\ref{eq:rs}) and (\ref{eq:lum}) with 
the standard analytical mass distribution,
one can compute the porosity, which is
defined as the product of the number density of sources and the bubble 
volume around each source:
\be
Q(z)  = \int_{M_{\rm min}}^\infty dM' \int_z^\infty dz' 
\, \frac{d^2 n}{dz' dM'}   \, V(\Eg,M',z,z'),
\label{eq:Q}
\ee
where $V(\Eg,M,z,z') \equiv 4 \pi r_s(\Eg,M',z,z')^3/3$ and
$r_s(\Eg, M',z,z')$ is the comoving radius at a redshift
$z$ of a shell from a source with total mass $M'$ that 
hosts an outflow at a redshift $z'$ with an energy input per
unit gas mass $\Eg$. Finally, $\frac{d^2 n}{dM' dz'}$ is the 
differential Press-Schechter comoving number density of objects forming 
as a function of mass and redshift calculated from:
\be
\frac{d n}{d{\rm ln} M} = \frac{\rho}{(2 \pi)^{1/2}M} \, \nu e^{-\nu^2/2} \,
\frac{d {\rm \ln}\sigma^2}{d {\rm ln} M}
\label{eq:ps}
\ee
where $\nu(M,z) \equiv \delta_c/[D(z) \sigma(M)],$ $D(z)$ is the linear
growth factor, $\sigma(M)^2$ is the variance associated 
with the mass-scale $M$, and $\delta_c \equiv 1.69.$
Note that although, strictly speaking, $\frac{d^2 n}{dM' dz'}$ accounts 
for both the creation of new sources and the destruction of older sources by
merging into larger objects (\eg Benson \etal 2005), it is sufficiently
close to the formation rate for the objects in which we are 
most interested here.

The porosity $Q$ can be thought of as the average number of
outflows impacting a random point in space, and it
depends on only two free parameters: $\Eg$ and the minimum mass,
$M_{\rm min}(z)$.   Here I assume for the fiducial case that 
efficient star formation occurs only in halos with
virial temperatures above $10^4 K$, because smaller objects are 
photevaporated after reionization, and
before reionization, gas cooling in these objects requires
$H_2$, which is an inefficient coolant
(Madau \etal 2001) that is easily dissociated
(Haiman \etal 1997; Ciardi \etal 2000).  
In our cosmology, this gives 
$M_{\rm min} = 2.4 \times 10^7 [(1+z)/10]^{-3/2} \msun,$ although 
I also consider variations in $M_{\rm min}$ in \S 3.2 below.

Carrying out similar integrals as in eq.\ (\ref{eq:Q}) we can 
compute ``porosity-weighted'' estimates of the properties
of typical sources.  For example the 
source mass that contributes most significantly to enrichment
can be estimated as
\be
\left< {\rm log}(M) \right>_Q =
Q^{-1} \int_{M_{\rm min}}^\infty dM' \int_z^\infty dz'  \
\frac{dn}{dz' dM'}   \, V(\Eg,M',z,z') \, {\rm log}(M').
\label{eq:lnM}
\ee
Similar averages can be used to compute the typical comoving bubble radius,
$\left<r_{b} \right>_Q$, the number density
of sources, $\left<\frac{dn}{d{\rm ln}M} \right>_Q$, and the source bias
$\left< b \right>_Q$, 
where $b(z,M) = 1+ \left[ \nu(z',M')^2-1 \right]/\delta_c$ (Mo \& White 1996).

\begin{figure}
\centerline{\psfig{figure=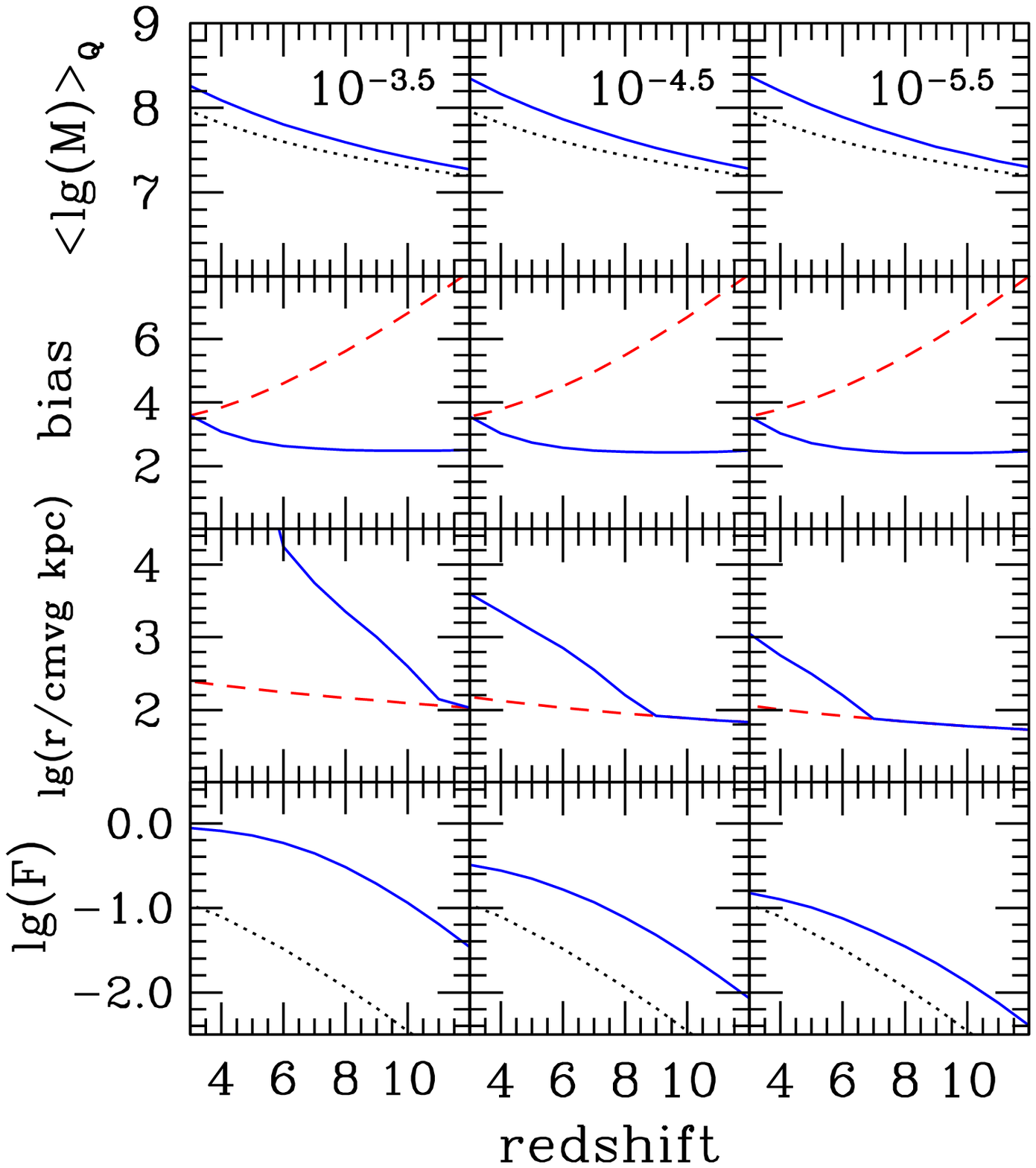,height=16cm}}
\caption{Properties of $T_{\rm vir} \geq 10^4$K enrichment for 
an energetic model with $\Eg = 10^{-3.5}$ (left column), the fiducial
model with $\Eg = 10^{-4.5}$ (center), and  a weak outflow model with
$\Eg = 10^{-5.5}$ (right).  {\em Top row:} Porosity-weighted  average
source mass as a function of redshift (solid) as compared to the
minimum mass (dotted).   {\em Second row:}
Porosity-weighed average bias $\left< b \right>_Q$ (dashed), and bias
normalized versus $z=3$, $\left< b \right>_Q \times D(z)/D(3)$ (solid)
which tracks the evolution of the correlation function as in
eq.\ (\protect\ref{eq:xibias}).
{\em Third Row:} Volume-weighed average comoving  outflow size, $\left<r_s
\right>_Q$ (dashed), and the typical comoving size of enriched region,
$\max\left\{r_{\rm overlap}, \left< r_s \right>_Q \right\}$ (solid),
which can be significantly larger due to overlapping sources.  {\em
Bottom Row:} Simple $1 - \exp(-Q)$ estimate of the filling factor (solid),
and 1/2 of the collapse fraction (dotted), an estimate of
the enriched gas falling back onto further generations of sources.}
\label{fig:T4}
\end{figure}

\section{Results}

\subsection{Variations in Energy Input}

The properties of three enrichment models with $T_{\rm vir} \geq 10^4$
K, and widely varying $\Eg$ values are shown in Figure 1.  In all
cases, the average source mass closely follows the minimum allowed
value.  The reason for this behavior can be seen from a simple
Sedov-Taylor estimate.  In this case the physical radius goes as $R
\propto (E/\rho)^{1/5},$ where $E$ is the energy of the blast  and
$\rho$ is the ambient density, and thus the
comoving volume goes as 
\be 
V_b \propto r_b^3 \propto M^{3/5} (1+z)^{6/5}.
\label{eq:ST}
\ee 
From eq.\ (\ref{eq:ps}) the comoving number density goes as $1/M$
(with only a small correction from the 
$\frac{d {\rm ln}\sigma^2}{d {\rm ln} M}$ term)
such that $Q \propto M^{-2/5}$.  Notice that this is a general result,
which follows from dimensional analysis.  The smallest sources
naturally dominate the enrichment process,  an effect that is only
amplified by complications such as the additional gravitational drag
in large halos (Scannapieco \etal 2002) or
the inefficiency of OB associations at driving winds from large
galaxies (Ferrara \etal 2002).  In fact only an
extremely strong increase in efficiency with mass ($\Eg \propto
M^n$, with $n \geq 2/3$) can alter this conclusion.

Similarly, the strong redshift scaling in eq.\ (\ref{eq:ST}) means
that for any given mass the earliest sources enrich most 
efficiently.
This is due to the simple fact that a fixed comoving
distance corresponds to a smaller physical  distance at higher
redshift, yet it has profound implications for the resulting spatial
distribution.  
In the second row of Fig.\ 1,  I plot the
porosity-averaged bias for each of the $\Eg$ models considered.  The
strong increase of $V_b$ at high-redshift means that the
relatively rare, high-$\nu$ sources contribute most strongly to $Q$.
This results in a high bias, as such rare sources are highly clustered
relative to the lower-$\nu$ peaks that collapse at lower redshifts
(\eg Mo \& White 1996).  Again this is a general result,
that arises from dimensional arguments, and the strong
clustering seen in this figure can only be amplified by IGM
transitions such as reionization, which remove lower-mass,
lower-redshift  (and hence lower-$\nu$) peaks (\eg 
Klypin \etal 1999).  Note that in general while bias increases with
redshift, the amplitude of the correlation function  is given by
\be
\xi\left(r,z,\left<b\right>_Q \right) = 
\left[ \left< b \right>_Q  D(z) \right]^2 \xi_0(r),
\label{eq:xibias}
\ee
where $\xi_0(r)$ is the matter correlation function, linearly
extrapolated to the present, and $D(z)$ decreases with redshift.
Thus $\xi(r,z,\left< b \right>_Q)$  
remains roughly constant with redshift.

The strong clustering of enriched regions must be also taken
into consideration when interpreting the typical sizes of enriched
regions.  In the third row of Figure 1, I plot 
$\left< r_s \right>_Q$, which again scales roughly as 
$\Eg^{1/5},$ and is always smaller than 300 comoving
kpc.  In order to estimate the sizes of typical
{\em measured regions}, however, we must also consider
the overlapping of such bubbles.
This can be estimated by considering the distance from 
the center of a typical source at which the product of the
 number density  of neighboring sources $n(r_{\rm overlap})$ and 
the volume around  each source  is equal to one.  In our formalism this gives
\be
\left[ 1+\xi \left(r_{\rm overlap},z,\left< b_L \right>_Q \right) \right] 
\times Q = 1,
\ee
where here it is more appropriate to calculate bias in  the Lagrangian
coordinate system that does not include the peculiar  motions of the
sources, as these motions were not included in our outflow model.
This means $\left< b_L \right>_Q = \left< b\right>_Q -1$ (\eg  Mo \&
White 1996).

\begin{figure}
\centerline{\psfig{figure=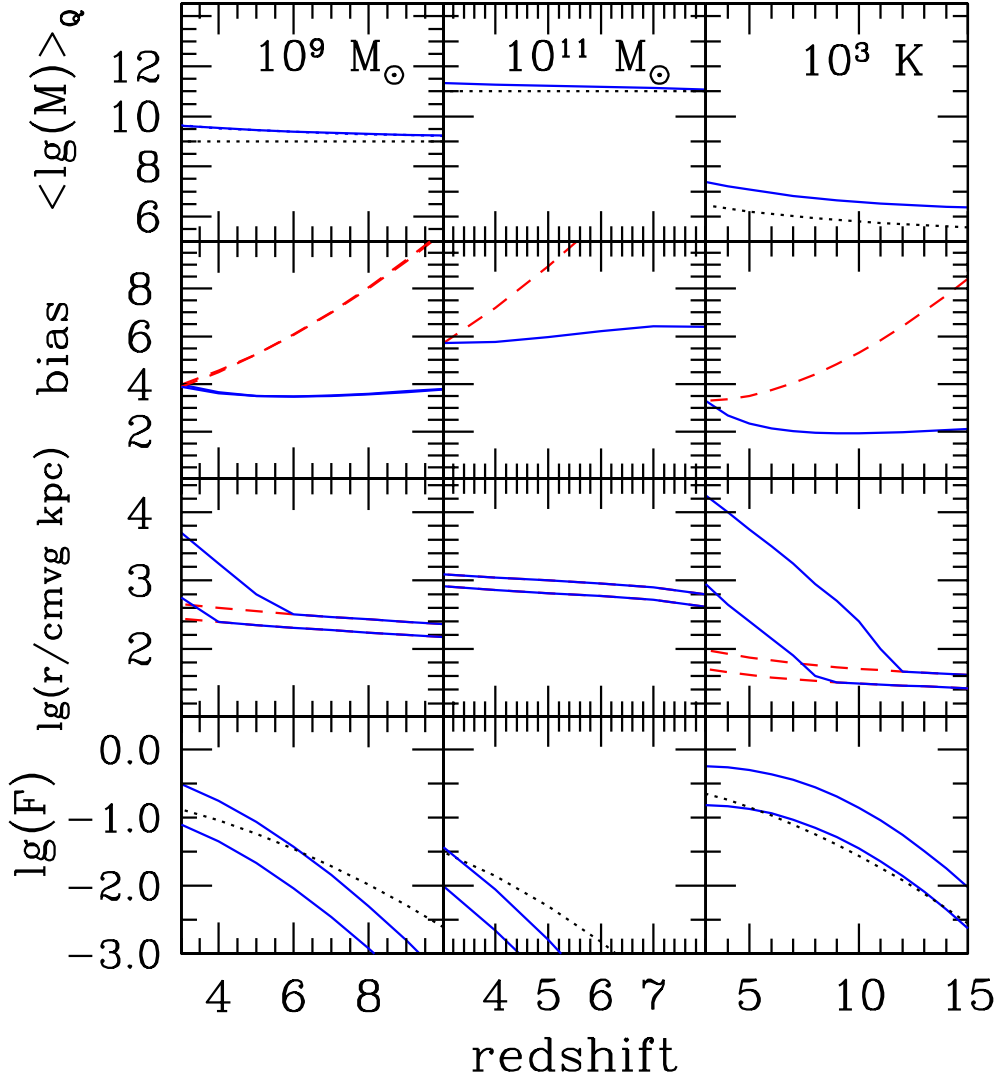,height=16cm}}
\caption{Properties of enrichment models as a function 
of minimum mass.  Each row is labeled by the assumed $M_{\rm min}$
value, while the curves in each row are as in Figure 1.  
In the left column $M_{\rm min} = 10^9 \msun,$ and I consider
two cases with $\Eg = 10^{-4.5}$ and $\Eg = 10^{-3.5}$.  In
the center panel, $M_{\rm min} = 10^{11} \msun$, and I take
even higher energies of $\Eg = 10^{-3.5}$ and $\Eg = 10^{-2.5}$,
which still give very low filling factors.  Finally the right column
shows the results of an extreme low-mass model in which 
$T_{\rm vir} \geq 10^3$K, and with low $\Eg$ values of
$10^{-4.5}$ and $10^{-5.5}$.}
\label{fig:mass}
\end{figure}

Despite the large range in $\Eg$ values considered, the overlap radius
is well above $\left< r_s \right>_Q$ at $z \leq 6$ for all three
models.  Thus the observed radii of enriched regions correspond to the
sizes of typical overlapping {\em groupings} of sources, rather than
the ejection radii from individual objects.   Similarly, the growth of
this scale with time is not caused by the expansion  of material from
a typical source, but rather is due to the formation of ever-larger
groupings of overlapping bubbles.  Indeed $r_{\rm overlap}$ increases
drastically at late times in all three models, greatly outpacing the
growth of the typical outflow as defined by $\left< r_s \right>_Q.$

The last row of Fig.\ 1 provides simple estimates of the filling
factor of enriched regions.  If the distribution of sources were
completely uncorrelated, this could be computed directly  from the
porosity as  $F = 1 - \exp(-Q),$ which corresponds to the solid lines
in these panels.  However, the strong overlapping between bubbles
means that the true filling factor is probably somewhat smaller than
this estimate.  Finally, the dotted lines in these panels
are 1/2 of the total collapse fraction, which is intended as a rough
estimate of the gas that would fall {\em back onto} 
new sources in a more detailed simulation.   This shows that infall
is not important in these models, except for perhaps a small
correction in the $\Eg= 10^{-5.5}$ case.

\subsection{Variations in Minimum Mass}

The second major parameter in our models is the minimum mass, which is
raised  to $10^9 \msun$ in the cases shown in the left column of
Figure 2, as might be caused by  the increase in the IGM temperature
following  reionization, for example.  This naturally raises $\left<
\log(M) \right>_Q$ by over an order of magnitude, but the bias is much
less affected, shifting from $\sim 3.5$ up to
$\sim 4.0$.   Thus although enrichment now occurs later, it is
dominated by sources  with similar $\nu$ values  as in the $T_{\rm
vir} \geq 10^4$ K case.   Also as in the fiducial models, $r_{\rm
overlap} > r_s$ in the observed redshift range, such that the scale of
enriched regions is  set by groupings of objects.  Interestingly, as
the filling factors are smaller in this case for the same values of
$\Eg$, raising the minimum mass actually {\em lowers} the sizes of
typical enriched regions.  However, there is probably a significant
infall correction in this model.

These effects are intensified in the extreme $10^{11} \msun$ model
shown in the center column of Figure 2.  In this case, the bias
rises to $\sim 5.5$, but the filling factors are so low that 
$r_{\rm overlap} < r_s$.  The observed bubble radii
are even smaller than in the $10^{9} \msun$ case, but now
one is actually looking around individual sources.
However, the very high energies and low filling factors
make this case unlikely.  Infall only makes this worse.

Finally, the right column of this figure shows the results of
an extreme low-mass model in which the minimum virial temperature
has been reduced to $10^3$ K.  Even in this case  
$\left< b \right>_Q$ is about 3.5 at $z=3.$  The radii of
enriched regions are again set by clustering, and are even larger 
than in the fiducial $T_{\rm vir} \geq 10^4$ K cases with the same 
choices of $\Eg.$

\section{Discussion}

It is clear that the simplified models described above gloss over many
of the detailed issues that are now beginning to be addressed 
numerically.  Nevertheless, this simplicity serves to highlight how
many counterintuitive observational trends can be understood  from
general arguments, which can be explored in more detail with
future simulations.

Thus, the seeming contradiction between widespread outflows 
from large $z \sim 3$ galaxies and the lack of evolution in  \CIV
number  densities is most likely related to the  $V \propto E^{3/5}$
scaling of outflows and the $1/M$ scaling  in the number density of
sources.  Likewise the strong clustering of  metal  line systems is
likely to be be reconciled with the efficient ejection of metals from
small $M \lesssim 10^{10.5} \msun$ galaxies (Tremoni \etal 2004)
because cosmological enrichment is dominated by early, rare sources,
that can expand to cover the same comoving volume in shorter
times.  Finally, as strong clustering results in a significant overlap
between sources, this may account for the large observed sizes  of
typical enriched regions,  which require extremely large ejection
energies to be explained by single sources (Kollmeier \etal 2003).
Although there is much more to be learned, it is clear that the
efficiency of lower mass, high-redshift  sources and the overlap
between bubbles should always be kept in mind when interpreting
measurements of the spatial distribution  of intergalactic metals.

\acknowledgements

I thank Andrea Ferrara, Pavel Kovtun,  Crystal Martin, 
Piero Madau, Michael Rauch, and an anonymous referee
for helpful comments and 
useful conversations.  This work was supported by the National 
Science Foundation under grant PHY99-07949.

\fontsize{10}{10pt}\selectfont

\end{document}